# New Insight on the Increasing Seismicity during Tenerife's 2004 Volcanic Reactivation

I. Domínguez Cerdeña [a], C. del Fresno [b], L. Rivera [c]

(a) Centro Geofísico de Canarias, Instituto Geográfico Nacional, c/La Marina, 20, 38001, S/C de Tenerife, Spain

(b) Observatorio Geofísico Central, Instituto Geográfico Nacional, c/Alfonso XII, 3 28014, Madrid, Spain

(c) Institut de Physique du Globe de Strasbourg (UMR7516 Université de Strasbourg, CNRS), 5, Rue René Descartes, F67084, Strasbourg, France.

## Abstract

Starting in April 2004, unusual seismic activity was observed in the interior of the island of Tenerife (Canary Islands, Spain) with much evidence pointing to a reawakening of volcanic activity. This seismicity is now analyzed with techniques unprecedented in previous studies of this crisis. The 200 earthquakes located onshore during 2004 and 2005 have been classified by cross-correlation, resulting in a small number of significant families. The application of a relative location algorithm (hypoDD) revealed important features about the spatial distribution of the earthquakes. The seismic catalogue has been enhanced with more than 800 additional events, detected only by the closest seismic station. These events were assigned to families by correlation and as a consequence their hypocentral location and magnitude were estimated by comparing them to the earthquakes of each family. The new catalogue obtained by these methods identifies two major seismogenic zones, one to the northwest and the other to the southwest of the Teide-Pico Viejo complex and having a separation of at least 10 km between them. These regions alternate their activity starting in January 2004, i.e., three months earlier than previously thought. We propose a simple model based on the results of this work which will also concur with all previous geophysical and geochemical studies of the 2004 crisis. The model proposes a single magma intrusion affecting the central part of the island with lateral dikes driven by the rifts to the northwest and southwest.

**Keywords**: Tenerife; Volcano-tectonic earthquakes; Reawakening of volcanic activity; Earthquake families; Relative location algorithm;

## 1. Introduction

During 2004 the island of Tenerife, Canary Islands (Spain) experienced a series of geophysical and geochemical changes which were considered evidence of a volcanic reactivation after almost 100 years of calm (Martí et al., 2009 and references therein). This crisis produced a significant social impact due to the high population of the island and the previous long repose periods in volcanic activity on Tenerife.

There was some controversy in the scientific community on the origin of these signals, with some scientists supporting a volcanic origin and the potential for the situation to culminate in an eruption (García et al., 2006; Gottsmann et al., 2006), while others disagreed with the possibility of volcanic reactivation (Carracedo and Troll, 2006; Carracedo et al., 2006). We must consider that this was the first volcanic reactivation ever detected in Tenerife using scientific instrumentation. This fact, together with the complex geology of the island, considerably increases the difficulties in interpreting the recorded signals. Although there were many measurements indicating the presence of a volcanic unrest episode, no eruptions have occurred to date.

The first sign of reactivation, or at least the one to trigger alarm, was the presence of onshore anomalous seismicity in Tenerife. Figure 1 shows the annual seismicity of Tenerife and its surroundings from 2001 to 2008. There is a seismogenic zone located southeast of Tenerife which is associated with tectonic activity (Mezcua et al. 1992). Although there seems to be increased activity from 2001 to 2008, this is an artifact due to improvements in the seismic network during these years. In April 2004 a new seismogenic zone appeared in the interior of the island located northwest of the Teide-Pico Viejo complex (first reported in the literature by García et al., 2006 and Gottsmann et al., 2006). This activity extended until 2005, with almost 200 located earthquakes of magnitudes from 1 to 3. Since 2006 the number of events detected in this region has decreased drastically.

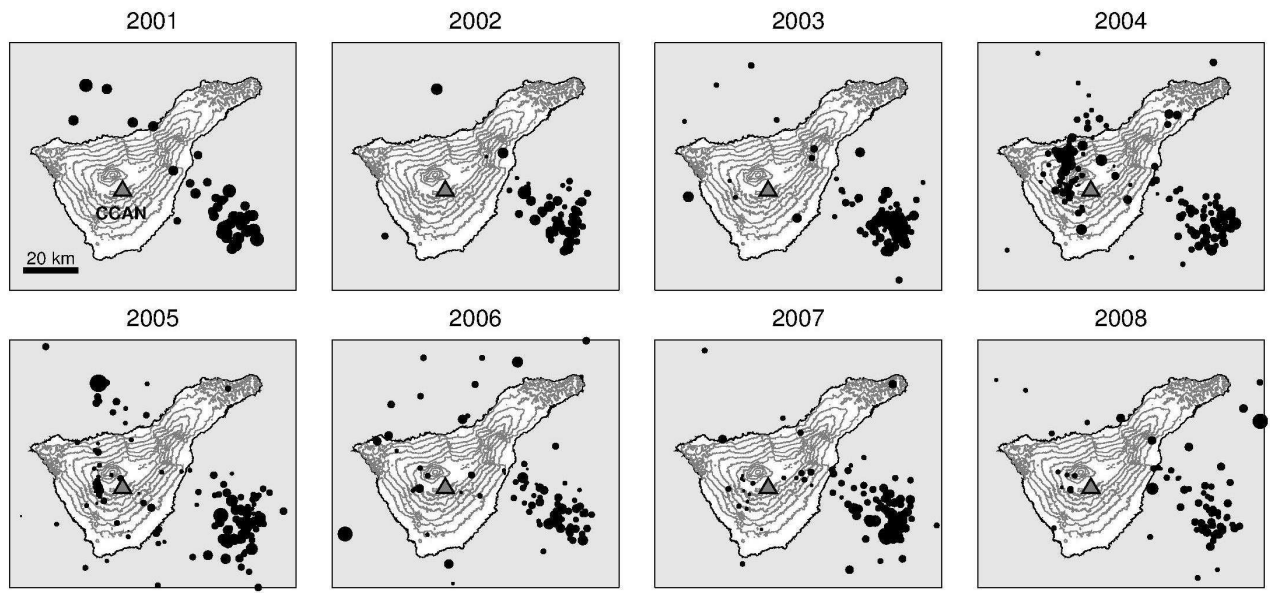

**Figure 1.** Annual seismicity on Tenerife and its vicinity from 2001 to 2008 according to the IGN catalogue. The size of the black dots is proportional to the magnitude of the earthquake. The maps include the position of the seismic station CCAN (gray triangle).

Further evidence of increasing seismic activity can be found in the number of events detected by the CCAN seismic station of Instituto Geográfico Nacional (IGN) on Tenerife (Fig. 1), which is the station best located for the study of this crisis and its instrumentation did not change during this period. The typical number of events per month detected by this station is around 50, but in April 2004 this number started to increase dramatically, reaching a maximum of 450 events during May 2004. From June 2004 the number decreased gradually back to the previous base level by the end of 2005. The number of detected events at CCAN from 2004 to 2005 was around 3600 as opposed to around 1200 which would correspond to a similar time period of relative calm, so there were a huge number of extra events that could be related to the seismo-volcanic activity.

There were other seismo-volcanic signals observed in that period as well, such as tremors and LP (García et al., 2006; Almendros et al., 2007), or low frequency seismic noise (Tárraga et al., 2006; Carniel et al., 2008a). At the time of this anomalous activity Gottsmann et al. (2006) detected microgravimetric anomalies with the maximum occurring at the northwest of Las Cañadas caldera and which seemed to migrate towards the western part of the caldera. Furthermore, there was an increase in the diffuse emission of carbon dioxide at Las Cañadas caldera (Galindo, 2005), an important increase in the $CO_2$ flux emission at Teide summit fumaroles (Pérez et al., 2005; Padilla et al. 2009), and presence of $SO_2$ and other magmatic gases at the fumaroles (Pérez et al., 2005; Pérez and Hernández, 2007; Melián et al., 2010). It is significant that those fumaroles were the only obvious sign of volcanic activity before 2004. A more detailed summary of the evidence of volcanic reactivation of 2004 can be found in Martí et al. (2009).

In this work we attempt to bring new insight to the body of knowledge on the seismo-volcanic crisis of 2004 in Tenerife: first, by offering some answers about the origin of this activity and, secondly, by increasing our understanding of the precursors for improved analysis of future reactivations. We also aim to improve some deficiencies of the IGN seismic catalogue of this region (Section 3). In order to obtain better information on the evolution of the seismicity, we have classified the earthquakes into different families (Section 4). We have also tried to enhance the seismic catalogue by improving the precision of hypocentral location for these events using a relative location algorithm and also by including some events of previously undefined location (Section 5 and 6). Finally, a new model is proposed to incorporate the results obtained in this work (Section 7).

## 2. Regional Setting

The Canary Islands are an archipelago of volcanic origin formed in the last 40 million years (Araña and Ortiz, 1991) and are located close to the northwest coast of Africa (Fig. 2a). This is the most active volcanic region of Spain, with at least 17 historic eruptions in Tenerife, Lanzarote and La Palma, over the last 500 years (Romero Ruiz 1991). The island of Tenerife was formed in two phases, starting with the growth of a basaltic shield in a period from between 12 and 3.3 million years ago (Ancochea et al., 1990) and continuing over the last 3.5 million years with the formation of a central structure including the Las Cañadas edifice (Martí et al., 1994). In the interior of the island the 16 km wide Las Cañadas caldera was formed by vertical collapses produced after intense explosive volcanic activity (Martí et al., 1997; Martí and Gudmundson, 2000) supported by recent geophysical works (Coppo et al., 2008; Gottsmann et al., 2008; Blanco-Montenegro et al., 2011). The Teide Pico Viejo complex is located in the northern part of Las Cañadas caldera (Fig. 2c) and gave origin to the explosive eruptions of the last several thousand years. Most of the effusive eruptions of the island originated in the NW and NE Rifts (Fig. 2c) and in the South Volcanic Zone. Both rifts are structural lineaments traced by aligned cones that extend from the central edifice towards the northwest and northeast of the island, while the South Volcanic Zone may be the result of the combined behavior of the two rifts (Geyer and Martí, 2010).

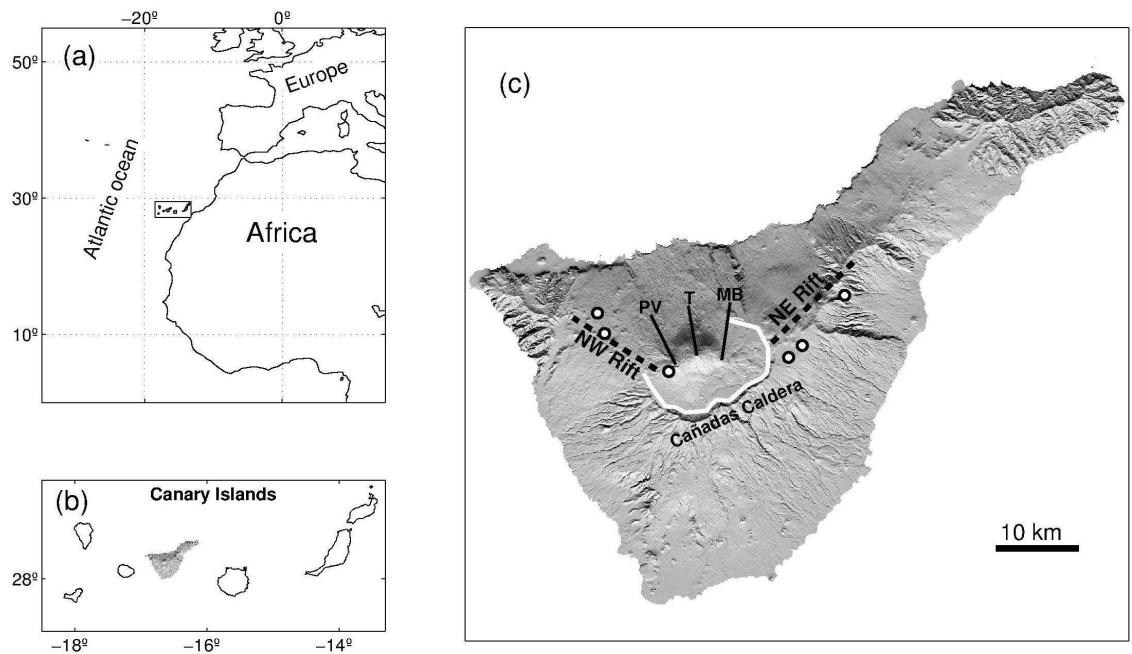

**Figure 2.** (a) Location of the Canary Islands. (b) The Canarian Archipelago with Tenerife highlighted. (c) Situation of the southern wall of the Cañadas Caldera (white line), the two Rifts (dashed black lines) and the three main structures in the interior of Cañadas Caldera, Teide (T), Pico-Viejo (PV) and Montaña Blanca (MB). Location of the vents of historic eruptions (white dots).

Many studies based on geological and geophysical observations have shown the volcanic structure of Tenerife with the presence of different magma chambers (e.g. Martí et al., 1994; Araña et al., 2000; Ablay and Martí, 2000; Martí and Gudmundson, 2000). They suggest the presence of a shallow magma chamber responsible for different phonolitic eruptions. Petrological studies of the phonolites from Teide indicate the presence of a magma chamber at a depth of 2 km (Ablay et al., 1998) while phonolites from the Montaña Blanca eruption suggest storage at sea level (Ablay et al., 1995). The study of the last eruption of Teide (Lavas Negras) shows a magma storage depth of 5±1 km below the summit (Andújar et al., 2010), thus of the order of 1 km below sea level. However, until now, there are no direct evidences of this chamber, not even with the seismic tomography study by García-Yeguas (2010). Different works have also suggested the existence of a deeper basaltic magma reservoir (e.g., Ablay et al., 1998).

The last volcanic eruption on Tenerife was observed in Chinyero in 1909. It was a basaltic strombolian eruption located on the NW rift of the island and, like most of the historical eruptions, was preceded by episodes of seismicity felt by inhabitants of the island. Historically, there have been only low explosivity basaltic eruptions, however, one of the last explosive eruptions was only 2000 years ago in Montaña Blanca (Ablay et al., 1995) which increases the potential hazard of the island.

Despite the historical and prehistorical volcanic activity and the numerous studies done in the Canary Islands, scientific monitoring of volcanic activity started only two decades ago. Current interest has increased substantially with the recent unrest episode (Araña et al., 1998; Blanco et al., 2006; Pérez and Hernández, 2008; López et al., 2008; Martí et al., 2009).

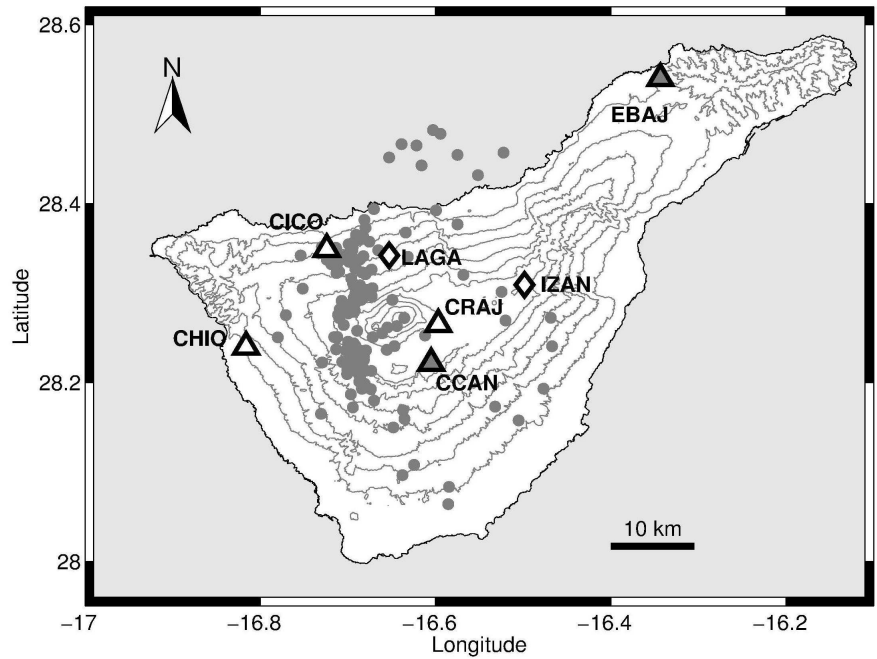

**Figure 3.** Seismicity on Tenerife (gray dots). We plot only those earthquakes in a radius of 25 km from Teide volcano. We also plot the seismic network located in Tenerife: permanent seismic stations installed before 2004 (gray triangles), permanent seismic stations installed from May 2004 (white triangles) and temporally deployed seismometers during the crisis (white diamonds).

## 3. Seismological data

The main data used in this work came from the National Seismic Network – Red Sísmica Nacional (RSN) and the seismic catalogue published by the IGN (www.ign.es). We used the waveforms of 200 events located in the interior of Tenerife Island during 2004 and 2005, recorded by stations located on the same island (Fig 3) and which have data for each earthquake from at least 3 and up to 7 seismic stations. Additionally, we use seismograms of 3600 non-located events from the CCAN station. First arrivals for these events have been manually picked.

There were obvious deficiencies in the original seismic catalogue. First, the accuracy of hypocentral locations was low due to the small number of seismic stations and difficulties in phase picking due to a low Signal to Noise Ratio (SNR). The second problem was a lack of homogeneity in the catalogue. The RSN of Tenerife was improved during the crisis in order to locate a higher number of events. Before April 2004 there were only two stations in Tenerife (CCAN and EBAJ; gray triangles in Fig. 3) and locations in the catalogue were obtained using data from two stations on the neighboring islands of La Gomera and Gran Canaria (EGOM and EOSO). Between May and July of 2004 three new permanent stations (CICO, CHIO and CRAJ; white triangles in Fig. 3) and two temporary seismometers were installed on Tenerife (LAGA and IZAN; diamonds in Fig. 3). The temporary stations had intermittent operating periods from June to November 2004. A catalogue generated from a changing Seismic Network cannot ensure its homogeneity and we found clear evidence of this problem in the completeness of the catalogue. By plotting only the monthly number of located earthquakes (Fig. 4a), we see a maximum of 37 events occurring in August of 2004. However, if we also plot the monthly (most of them previously un-located) events detected by CCAN (Fig. 4b) we appreciate that the maximum of 450 events occurred in May 2004. Thus, there seems to be a depletion of seismicity before July 2004 that should be included to complete the catalogue.

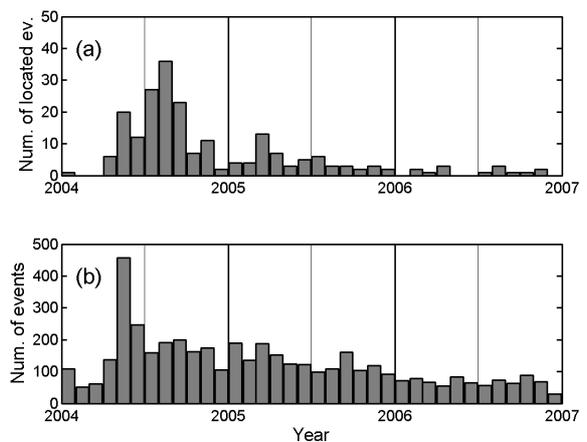

**Figure 4.** (a) Histogram of the number of monthly earthquakes from the IGN catalogue located in Tenerife (those shown in Fig. 3) from 2004 to 2007. (b) Histogram of the monthly events detected by the seismic station CCAN during the same period.

The seismometers used by RSN are basically of two types. The short period stations (CCAN, CICO and CHIO) use the 1 Hz Kinemetrics SS1-Ranger with a sampling rate of 50Hz and only the vertical component. The other permanent seismic stations (EOSO, EGOM, EBAJ and CRAJ) use a three-component Broad-Band (BB) sensor, i.e. CMG-3T from Guralp with a sampling rate of 100Hz. The SP stations use Kinemetrics TH-11 telemetry, while the BB stations is digitize by a 24-bit Trident and Cignus VSAT (Nanometrics). Temporary stations set up during the seismic crisis (LAGA and IZAN) used BB sensors CMG-3T with a sampling rate of 100Hz, and a Geotech DL 24-bit digitizer. Further details of each station can be checked in the IGN webpage (www.ign.es). Waveforms from temporary stations are not used for correlation due to their poor quality and data from EOSO station was not considered due to its large distance to the seismic zones. The only data used from these stations is the arrival times of the seismic catalogue.

## 4. Earthquake classification

As a first step in this study the 200 located earthquakes occurring below the island of Tenerife during 2004 and 2005 (those from the IGN catalogue) were classified into different families by waveform cross-correlation. This method has been successfully applied in classifying the seismic signals of many volcanoes, including both volcano-tectonic seismicity (Okada et al., 1981; Lahr et al., 1994; Zamora-Camacho et al., 2007; Umakoshi et al., 2008; Carmona et al., 2010) and Long-Period events (Stephens and Chouet, 2001; Green and Neuberg, 2006, Varley et al., 2010). Each family obtained by this method should include events produced by a very similar focal mechanism and that are tightly grouped within a small volume (Okada et al., 1981). Ray trajectories from source to station of two earthquakes of the same family and recorded in the same station should be very similar (within less than one wavelength).

The temporal range for the correlation was chosen on the base of two considerations to ensure that the whole seismic signal of the earthquake was included: First, slight errors in arrival time are possible due to noise and manual picking. Second, the earthquake duration varied from one station to another depending on hypocentral distances. For that reason the time window of the seismic signal used was chosen between three seconds before the first arrival (obtained from the catalogue) and three times the time delay between P and S waves of the earthquake located closest to the centroid of the whole cluster.

Another difficulty in cross-correlation analysis is the presence of noise. Because we are working on an island there is a high microseismic noise at low frequencies (Braun et al. 1996; Almendros et al., 2000). To address this, a band-pass filter of between 2 and 8 Hz was used, as this is the range that contains most of the energy of these quakes. Additionally, because the noise level of each trace depends on the station, we have assessed a weight to each waveform as the logarithm of the SNR.

Finally, the classification was performed using seismic signals from 6 stations, which gives high reliability to the results. Only the vertical component seismic data is used for correlation, because the major percentage of seismograms was recorded by the one component seismic stations. The correlation matrix for each seismic station was obtained computing the correlation and time lag between each pair of earthquakes. A definitive single correlation matrix was computed as a weighted linear combination of the different matrices with the weights estimated beforehand. After a hierarchic analysis we chose a threshold level for a correlation factor of 0.6 to create the families. This value may be considered low compared to other earthquakes classification studies which use threshold values as high as 0.9 for tectonic (e.g., Saccorotti et al., 2002; Carmona et al., 2009) or volcanic seismicity (e.g., Carmona et al., 2010), however, these works used shorter correlation windows, e.g., 1 second (Carmona et al., 2010), while we are using windows larger than 10 seconds. Moreover, our correlation matrix is the combination of data from 6 seismic stations which increase the strength of the result but decrease the value of correlations. Many other previous works have used similar values, e.g., 0.7 (Green and Neuberg, 2006), 0.68 (Stephens and Chouet, 2001) and 0.6 (Umakoshi et al., 2008; Varley et al., 2010), all of them used correlation windows between 8 and 10 seconds.

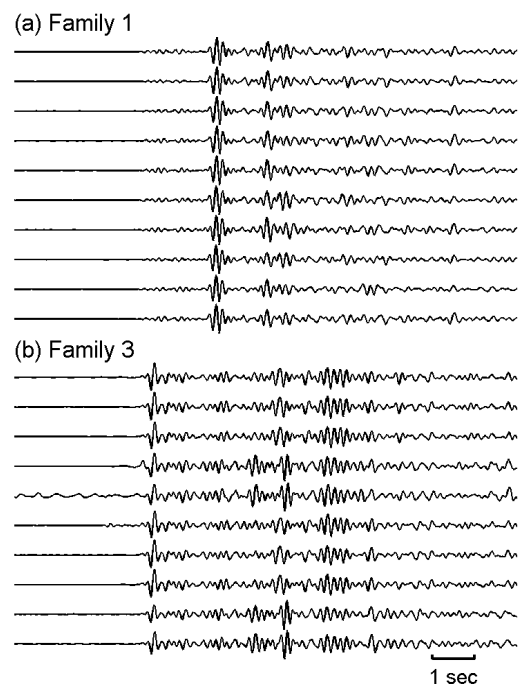

**Figure 5.** Seismograms of some events belonging to families 1 (a) and 3 (b) recorded in CCAN station. Total length of seismograms is 12 s and seismic signals are normalized.

The classification resulted in six major families including almost 60% of the total events. The first family contains 53 earthquakes but only four families include more than 10 events and the number of events per family decreases to two for family 20. Only 36 earthquakes remained not grouped. In Figure 5 we compare 10 individual events belonging to families 1 and 3. Events belonging to the same family are very similar all along the duration of the earthquake even though the time lag between the first and last event of each family can be of some months.

Figure 6 shows the hypocentral distribution of the six major families. We can distinguish two seismogenetic zones. The first (from now on called cluster N) is located in the northwest part of the island and it includes families 1, 2 and 4. The second (cluster S) is to the south of cluster N and contains events from families 3, 5 and 6. In general, earthquakes belonging to the same family are located close to each other, as was expected, however, we found some suspicious exceptions and some of the families are not as grouped as they should be (in particular, family 1). In most cases the catalogue has no depth computed for the events, usually due to the low number of seismic stations.

## 5. Seismic catalogue enhancement

Because, as we have already pointed out, there were deficiencies in the hypocentral locations of earthquakes in the IGN seismic catalogue of Tenerife (Section 3), we improved them by using a relative location algorithm. There are many possible sources of error in determining absolute hypocenter location, such as arrival time readings, seismic network geometry and/or uncertainties in crustal structure (Pavlis 1986). Relative location methods could minimize the effects of these uncertainties in the structure model (e.g., Poupinet et al., 1984).

Another deficiency found in the IGN seismic catalogue of the 2004-2005 crisis was the small number of located earthquakes in Tenerife (470) compared with the large number (3600) of detected events at CCAN seismic station. This station usually detects about 50 events per month, which was the background level before the crisis started. Thus, it should have recorded around 1200 events in 2 years, ergo; there were 2400 anomalous events that could be associated to volcanic activity. In order to obtain some information on the location of these events we have tried to assign them into the previously obtained families (Section 4). Once grouped, the hypocenter of each earthquake should be close to the centroid of the corresponding family.

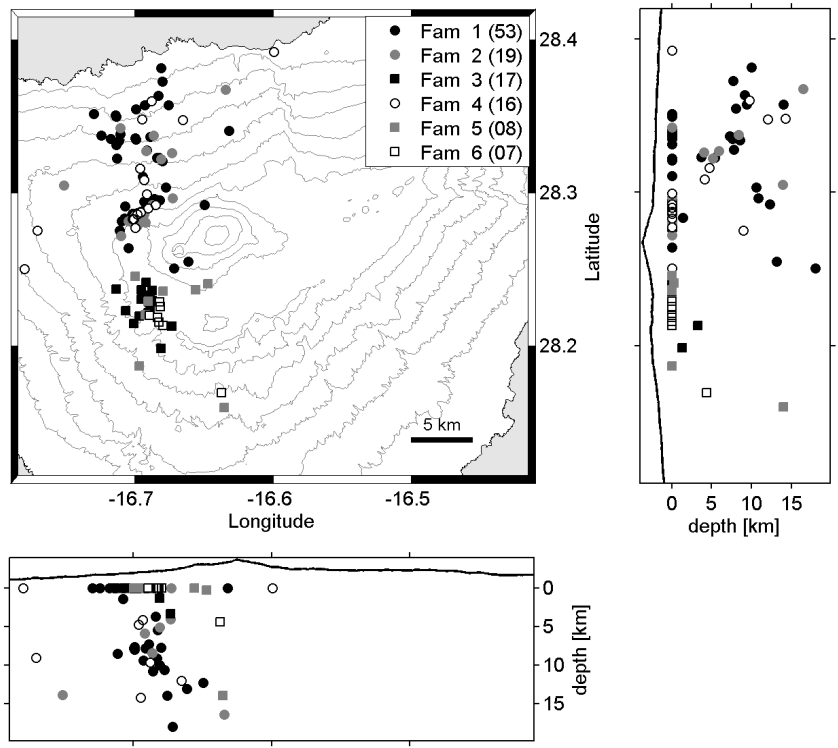

**Figure 6.** Catalogue location for the events belonging to the six most populated families. Depth of the events is shown in a vertical cross-section in N-S direction (upper right panel) and the E-W direction (bottom panel).

We also took advantage of the classification and assignment of events to homogenize the catalogue by recalculating the magnitude of each shock. As previously suggested, magnitudes computed for the seismic catalogue may have been affected by changes in the seismic network during 2004. The used methodology also allows us to assign magnitudes to the new events detected at the CCAN station. This is not the first time magnitudes have been obtained using only one station; for example, De la Cruz-Reyna et al. (2008) calculated the coda magnitude at Popocatepetl.

*5.1. Hypocentral relative relocation*

We have used HypoDD algorithm (Waldhauser and Ellsworth, 2000) to perform the relocation. This algorithm uses double difference equations to improve location precision by removing the effects of the non modeled velocity structure. If the distances between earthquakes are small compared to the epicentral distances, the double difference method is insensitive the heterogeneities of the velocity model. This is an interesting feature as, in Tenerife, crustal velocity has variations in the order of 40% along few kilometers (García Yeguas, 2010). At the same time, this method has provides an important improvement with respect to other relative location methods like JHD (see Waldhauser and Ellsworth, 2000).

The HypoDD algorithm combines the travel time differences from each phase (P or S) obtained from manual picking and/or computed by cross-correlation. This combination of data is one of the main advantages of the method because the use of cross-correlation data considerably increases the accuracy of the location. Travel time differences from manual picking have been obtained from the IGN catalogue. Their values have been computed automatically from all possible combinations of events. The precision that can be achieved is obviously never better than the sampling period (0.02s for SP stations and 0.01s for BB stations). When we compared the seismic signals of some earthquakes from the same family whose phases should be easily recognizable, we realized that, in many cases, manual picking had an even lower accuracy probably due to the noise level. Travel time differences from cross-correlation have been computed automatically and corrected manually in the noisy cases. We took into account only differences where correlation is larger than 0.6. In this case, the correlation was only possible between events of the same family or similar families. This includes, on the one hand, all events from North cluster (e.g. families 1, 2 and 4), and on the other hand, all events from South cluster (e.g. families 3, 5 and 6). The accuracy in this case was much higher than for catalogue data, achieving a precision of 10% of the temporal sampling (1-2 msec) when the seismic signals had a sufficient coherency (Poupinet et al. 1985). Such a precision is an important improvement, mostly because the manual picking of the S phases was not clear in most of the cases.

**Table 1.** Velocity model from Del Fresno et al. (2008). With $Z$ the bottom depth below sea level of each layer, $V_p$ the P- wave velocity and $V_s$ the S-wave velocity

| $Z$ [km] | $V_p$ [km/s] | $V_p/V_s$ |
|---|---|---|
| 0.3 | 4.7 | 1.65 |
| 6.3 | 5.2 | 1.79 |
| 8.3 | 6.5 | 1.77 |
| 10.3 | 7.3 | 1.75 |
| 12.3 | 7.8 | 1.73 |
| $\infty$ | 8.2 | 1.76 |

HypoDD minimizes residuals between observed and computed travel time differences with an iterative method. After each iteration it updates the hypocentral location and corrects the weights of the data depending on certain parameters. Initial locations used in the process were those from the catalogue. The velocity model used was obtained for Tenerife by Del Fresno et al. (2009) and computed from the detailed analysis of the seismic events with higher SNR (Table 1).

The method allows for an *a priori* weighting of the data which was done individually for each time lag between each pair of phases. In the case of the catalogue data, each phase is assigned with four possible weight values, i.e., 1.0, 0.2, 0.1 or 0.05 depending on the pick quality (in the case of the IGN catalogue this quality is assigned manually in four levels). As we have explained, we expected a low precision in the catalogue picking of the S phase and thus we reduced its weight to 0.3. The weight of the cross-correlation data was given by the squared coherency. Another key parameter is the weight given to each data type, as we expected more accuracy on the differential time computed by correlation than from the catalogue data, the weight for that data was set to 0.5, while the weight of the cross-correlation data was 1.0. The re-weighting parameters, which correct the weights after each iteration, were set to a cutoff for event separation larger than 5 km and of residuals of 5σ (more details in Waldhauser and Ellsworth, 2000).

The total data used included up to 66,000 differential times, 74% of which were catalogue data and the remainder cross-correlation data. In both cases there were almost balanced numbers of P and S phases. For some events the amount of data available was very large, with up to 500 catalogue differences and 300 cross-correlation differences. In the case of catalogue data we have taken into account only the differential times between events closer than 10 km to ensure the fulfillment of the conditions for hypoDD algorithm. Since the cross-correlation data was taken between events from similar families it was not necessary to apply this cutoff.

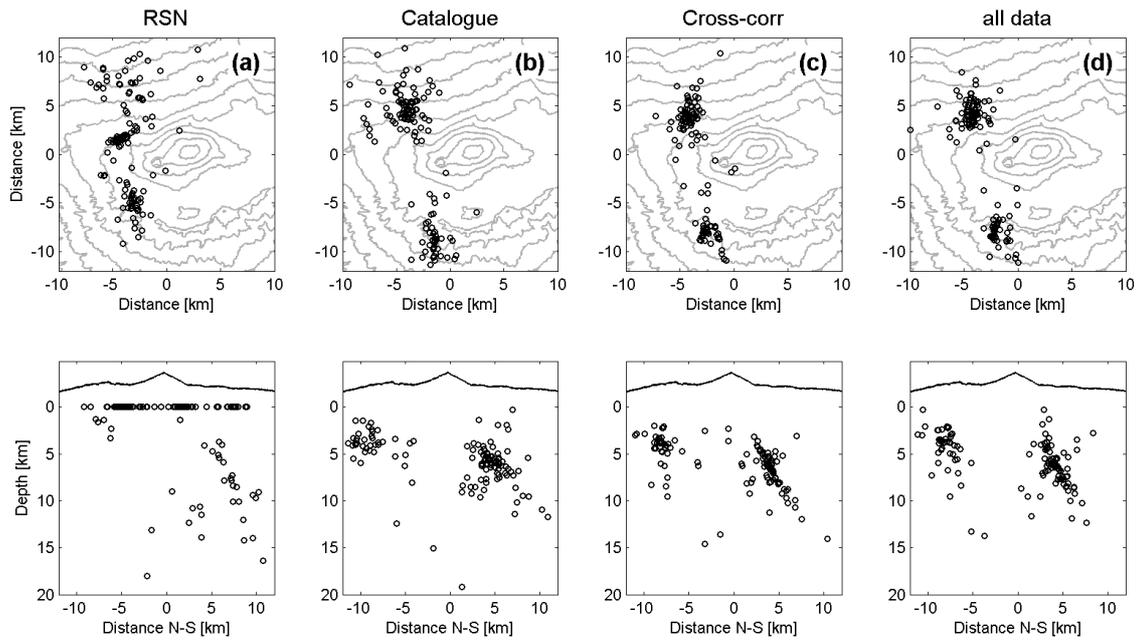

**Figure 7.** Results from the hypoDD relocation. Comparison between original catalogue location (a), relocation obtained only with catalogue data (b), relocation only with cross-correlation data (c) and relocation with the combination of all data (d). Upper panels show epicentral locations and lower panels show vertical cross-sections in N-S direction.

Figure 7 compares the original locations from the IGN seismic catalogue with the relocations obtained using catalogue data only, cross-correlation data only and a combination of both. These results were obtained after 10 iterations, which ensure the convergence of the method. In the case of relocation using all the data, the mean residual time obtained was approximately 500 ms for the catalogue data and 50 ms for the cross-correlation data. More than 70% of the events were relocated as well as about 88% of those from the 6 major families; however, some earthquakes were not relocated as they were rejected when their depth was negative.

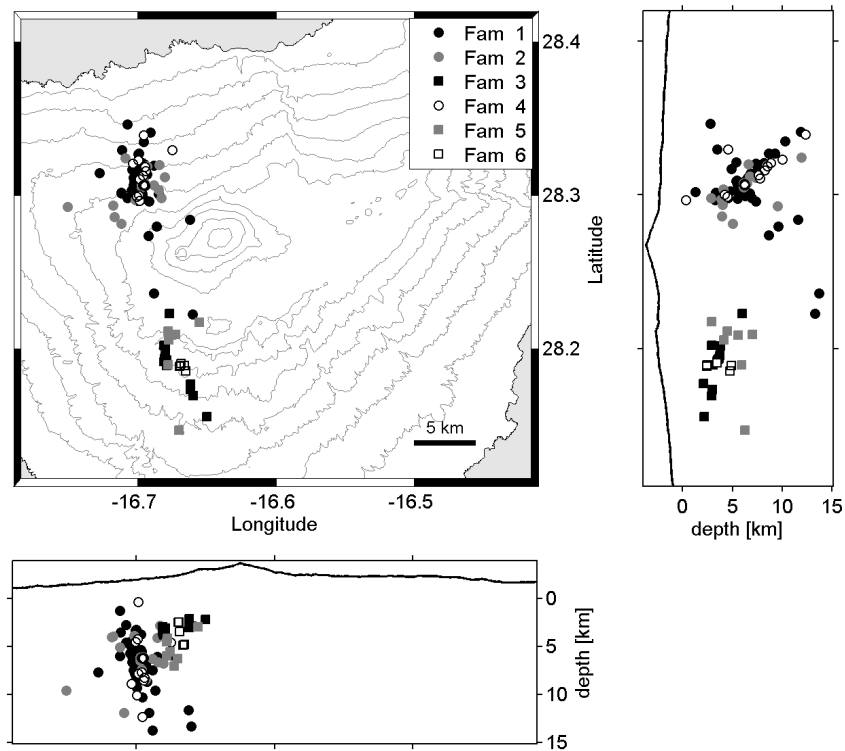

**Figure 8.** Relocation for the events belonging to the six most populated families. Depth of the events is shown for vertical cross-sections in N-S direction (upper right panel) and E-W direction (bottom panel).

There were two improvements on results with respect to the original locations: First, both clusters (North and South) see their events getting closer, and second, there is much more information about the depth of the earthquakes. Location distributions obtained from different data show a similar behavior; however, the results obtained with catalogue data are more spread than those computed with cross-correlation data and all data, while these two are almost equally distributed. This comparison gives some consistency to the results obtained by two different sets of data, and also shows the need of the correlation data which substantially improves the localizations. Waldhauser and Ellsworth (2000) show really accurate results, however, they have data from a huge number of seismic stations using more than 40 phases per event while in our case the maximum number was 12 phases. Furthermore, their data has a higher ratio between distance from events to the stations and extension of the seismic cluster than the data used in this work.

Considering the relocation done with all data, the first characteristic we can observe in the relocated distribution of events is the increasing distance between cluster N and S when compared to the IGN locations. The centroid of cluster N is placed farther north while the centroid of cluster S is located farther south from its original position. All families show their events more closely grouped (Fig. 8), and in some cases the horizontal separation between earthquakes is lower than 1 km (family 6). Another conspicuous feature is the depth distribution of the families, with cluster S located between 2 and 7 km below sea level, and cluster N between 3 and 12 km below sea level. Both clusters seem to be aligned in a trend pointing to N and towards larger depths, with the exception of a few outliers. However, the inclination of the northern cluster is much larger and the alignment is straighter.

*5.2. Assigning events detected at CCAN station*

First, we have computed the average family waveform of the first 20 families from the stack of all the events of each family. Each seismogram of the 3600 events was filtered and cropped in the same way as the classified events (see more details in Section 4) and then compared to those 20 mean traces by cross-correlation. We assigned each event to the family with the best correlation whenever the coherence exceeded a threshold value of 0.5. To check this method we compared it with the catalogue events that were already classified and were able to retrieve the families with more than 90% success. As it happened with classification of catalogue earthquakes, seismograms of the same family look very similar (Fig 9), revealing they all have similar location and focal mechanism. This fact is observed even several seconds after the first arrival.

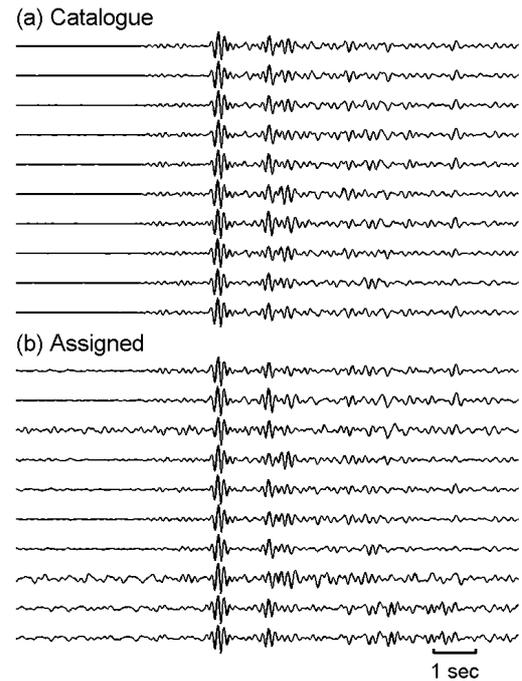

**Figure 9.** Seismograms at CCAN station of some earthquakes of IGN catalogue belonging to family 1 (a) and some of new events assigned to the same family (b). Total length of seismograms is 12 s and seismic signals are normalized.

More than 800 new events were assigned to the existing families. Results showed that almost 67% of the events belong to the South families while the remaining 33% correspond to the North families. The most populated is family 3 with more than 200 events, followed by family 1 with 150 events. Many of these events were assigned to the major families (Fig. 6); however, some of the low populated families grew considerably, like family 12, which previously had only 3 located earthquakes but contained more than 60 events after allocation of the CCAN events.

In order to make the interpretation easier, we analyzed the time evolution of the North and South clusters. Figure 10 shows time versus level of seismic signal (logarithm of RMS) of the IGN catalogue events and of the total number of events (catalogue + new events assigned). The detection level of the IGN catalogue events (Fig. 10a) seems to decrease with time, from April to August of 2004, and there is a void of events below log(RMS)=1. The existence of this void was already suggested in Section 3 (see Fig. 4). The catalogue events are mostly located in the North at the beginning of the crisis (April 2004) with few events in the South. After September 2004 all seismic activity seems to have migrated to the South. In the case of the total events (Fig. 10b) the detection level of the station remains stable at all times for values slightly lower than log(RMS)=0 and an additional characteristic not seen before is observed: The seismic activity appeared on January 2004 in the South, followed by a period of relative calm of a couple of months, and then from April to September it alternated between South and North clusters. Finally the seismic activity detected since September 2004 is mostly in the South cluster.

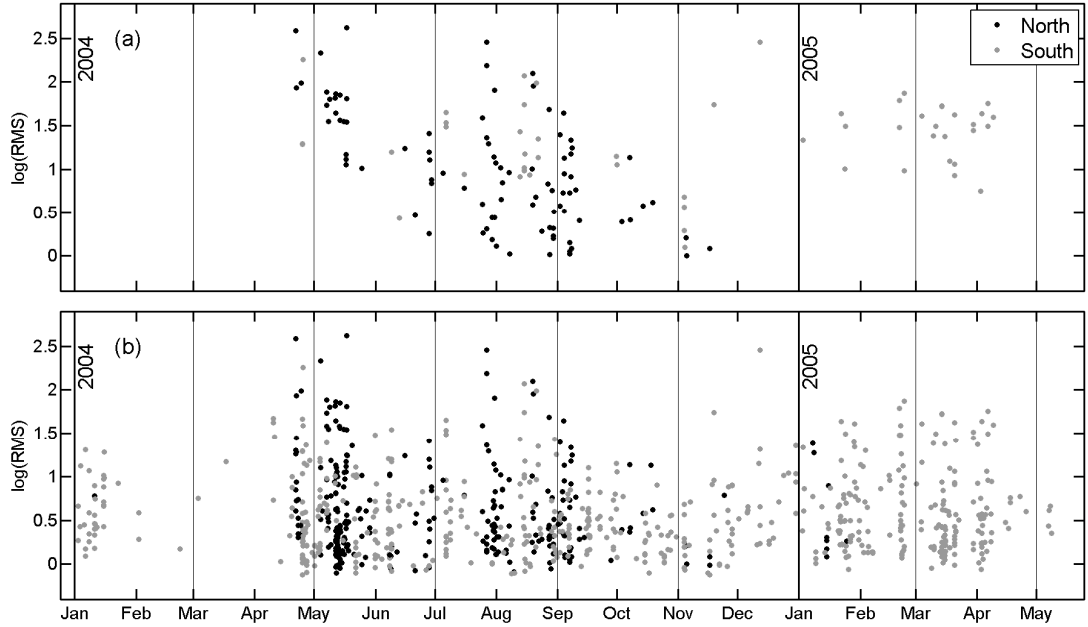

**Figure 10.** Temporal distribution of the log(*RMS*) of the events belonging to North families (black dots) and South families (gray dots). Comparison between earthquakes from the IGN seismic catalogue (a) and from the improved catalogue (b).

*5.3. Magnitude estimation*

The magnitude of the earthquakes located in the Canary Islands by the Spanish RSN is computed by the formula

$$m = log\left(\frac{A}{T}\right) + 1.17\, log(d) + 0.0012\, d + 0.67 \quad (1)$$

where $A$ is the amplitude in μm of the S phase, $T$ the period in sec. of the wave and $d$ the hypocentral distance from the earthquake to the seismometer in km. The magnitude is calculated at each seismic station and then averaged to obtain a single value that minimizes the influence of the orientation of the focal mechanism and of the different local effects. The displacement $u$ produced by an earthquake in a seismometer can be expressed as:

$$u \sim M_0 R^c_{\theta\varphi} f(d) \quad (2)$$

with $M_0$ the seismic moment which is proportional to the magnitude, $R^c_{\theta\varphi}$ the radiation pattern function, which depends on the focal mechanism and orientation from the hypocenter to the sensor and $f(d)$ a function depending on the distance and raypath. Earthquakes belonging to the same family are expected to have the same focal mechanism and to be located approximately in the same area. Thus, we assume that the amplitude of the observed ground motion only depends on the seismic moment, ergo, only depends on the magnitude. With this argument we can propose the next hypothesis: the magnitude $m^n_{i,j}$ of the event $i$ belonging to family $j$ measured from the seismic station $n$ can be approximated to the logarithm of the RMS of the seismic signal of this event at this station plus a constant $C^n_j$ depending on the family and on the seismic station.

$$m^n_{i,j} = \log\left(RMS^n_{i,j}\right) + C^n_j \quad (3)$$

This equation can be considered analogue to Eq. 1, but only valid to obtain magnitudes of events belonging to a single family by means of signal comparison. The RMS of the seismic signal is equivalent to the mean velocity measured at the seismometer and could replace the maximum velocity of Eq. 1 (*A/T*). The constant $C^n_j$ includes the information of the focal mechanism and of the between the catalogue magnitude and log(RMS) for each event of a family the result should be a constant (Eq. 3). However, when we did this for events of family 1 (Fig. 11) the result was unexpected. Events before June 2004 showed a value below zero and since July the events showed a value around 0.5. The reason for such behavior could be the change produced in the seismic network. Before June the earthquakes were analyzed with only 3 seismic stations resulting in a poor magnitude determination. After that time, new seismic stations were installed and there was an improvement in the magnitude estimation. Therefore, we will apply Eq. 3 to every event in order to homogenize the catalogue.

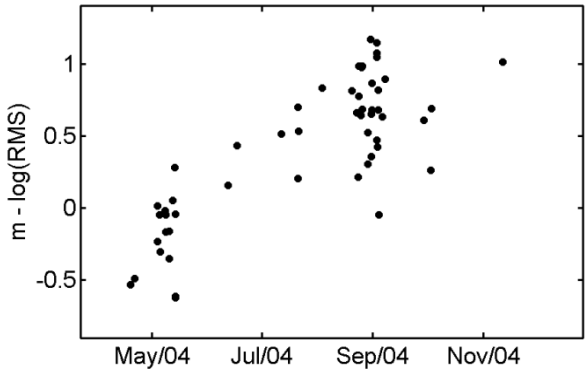

**Figure 11.** Time evolution of the difference between magnitude and log(*RMS*) of the seismic signal for the events corresponding to family 1.

The constant $C_j^n$ was obtained by inverting Eq. 3 for the events whose magnitude is well computed (those with index *l*) and averaging:

$$C_j^n = \frac{1}{N_j}\sum_{l=1}^{N_j} m_{l,j}^n - \log\left(RMS_{l,j}^n\right) \qquad (4)$$

We considered the best magnitudes calculated to those computed using at least 6 seismic stations. This constant has values close to +0.2 for the North families and to –0.4 for the South families. The detection level for both clusters is similar (Fig 10), hence we inferred that CCAN is more sensitive to events from the South cluster than those from the North cluster as can be seen in the histograms of the number of events (Fig. 12).

Although this is not a very robust technique to estimate the magnitude, it was the best approach we could apply to data from one single station and taking advantage of the waveform classification. Even if the magnitudes are not exact, the relation between events of the same family should be accurate.

## 6. Magnitude of completeness of the new catalogue and Gutenberg-Richter b parameter

One method to analyze volcanic seismicity is to study the magnitude distribution according to the Guttenberg and Richter law (1944), which relates it with the frequency of occurrence:

$$\log(N) = a - bM \qquad (5)$$

with *N* the accumulative number of earthquakes of magnitude larger than or equal to *M*, and *a* and *b* are empirical parameters that need to be fitted on each region. This relation is valid only when considering magnitudes larger than the magnitude of completeness of the catalogue ($M_c$). The magnitude of completeness is defined as the lowest magnitude at which the total number of events is detected at a given region and in a given period of time (Rydelek and Sacks, 1989; Wiemer and Wyss, 2000).

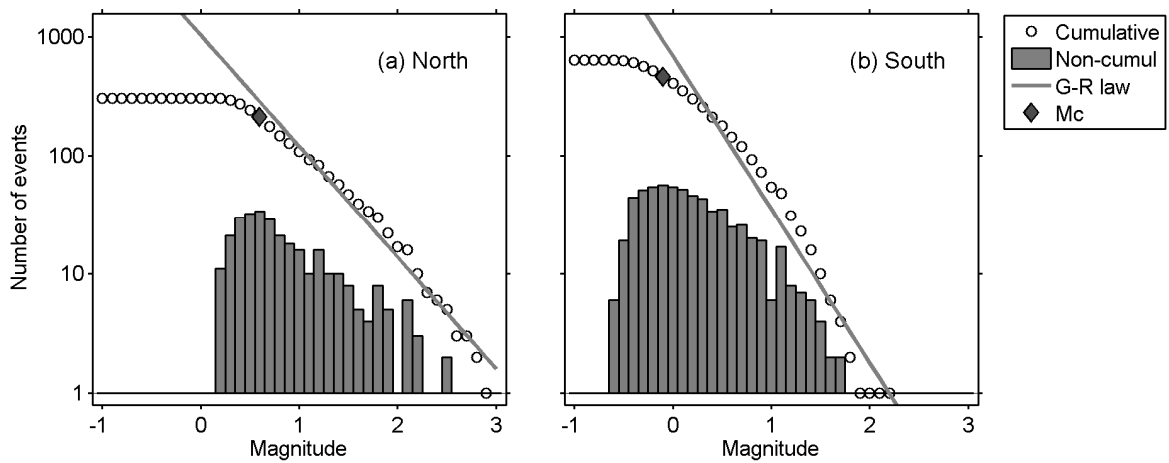

**Figure 12.** Gutenberg-Richter curves for the events belonging to North families (a) and South families (b). Open dots represent the cumulative number of events while histograms show the non-cumulative distribution of events in magnitudes. Gray diamonds show the completeness magnitude $M_c$ in each case. The gray lines represent the best fit to the Gutenberg-Richter law.

Parameter *b* of the Gutenberg-Richter curve gives information on the nature of the medium where the earthquakes are originated. The b-value is inversely related to the stress applied and directly related to the heterogeneity of the material (Mogi, 1967; Scholz, 1968). Tectonic seismicity is usually associated with b-values of the order of 1, however, in many volcanic scenarios, where the seismicity surrounds magma chambers or is associated with the volcanic activity, this parameter reaches higher values. Some studies have obtained b-values even larger than 2 in volcanic regions (e.g., Wiemer and Mc Nutt, 1997; Wiemer and Wyss, 1997; Wyss et al., 1997; Novelo-Casanova et al., 2006). Moreover, volcanic areas have shown considerable variations in the spatial distribution of the b-values with changes from 1 to even more than 2 at a distance of few km (e.g., Wyss et al., 2001; Farrel et al., 2009)

Since we have detected two different and clearly independent seismogenetic zones, we have studied the $M_c$ and b-values for each region separately. Figure 12a and 12b show Gutenberg-Richter curves for the North and South events respectively and histograms of non-cumulative magnitude distribution for both clusters, showing lower values for the Southern events. To compute $M_c$ we have used the maximum curvature method (MAXC by Wiemer and Wyss, 2000; Woessner and Wiemer, 2005) which simply chooses $M_c$ as the maximum value of the first derivative of the frequency-magnitude curve. This value matches the highest point of non-cumulative distribution. In our case (gray diamonds in Fig. 12) we found values of $M_c$ = –0.1 for South events and $M_c$ =+0.6 for North events. The fraction of earthquakes with magnitudes larger than $M_c$ is 62% for the North zone and 70% for the South cluster.

We have followed two techniques to ensure the completeness obtained by the MAXC method. Both assume that detection of low magnitude events is easier at night due to lower noise produced by human activity, a problem that was highlighted for Tenerife IGN catalogue by Carniel et al. (2008b). The first technique is the completeness test proposed by Rydelek and Sacks (1989). It studies the preference for detecting earthquakes as relating to time of the day. With this objective, a phasor is assigned to each event oriented in a 24 hours clock, towards the time of the day when it is detected. Each phasor is added at the end of the previous one for all earthquakes included in a magnitude interval. For complete catalogues, the phasor sum should be a random walk around zero. However, if the catalogue is incomplete the phasor sum will point to night hours in the 24-h clock. To decide whether the catalogue is complete or not, the phasor sum is compared to the expected random walk of a Poisson distribution of earthquakes. In other words, the phasor sum should not exceed a radius of $1.73\sqrt{N}$ which corresponds to the 95% confidence level for a random walk for a sample of $N$ earthquakes (Rydelek and Sacks, 1989; Centamore et al., 1999). We have applied this test to earthquakes in three different magnitude ranges: lower than $M_c$, between $M_c$ and $M_c+0.6$, and for larger than $M_c+0.6$. Results show similar behavior for North (Fig. 13a) and South data (Fig. 13c), with the low magnitude events clearly out of the confidence circumference and oriented towards night hours. The other two magnitude intervals are clearly inside the circle proving the completeness of the catalogue for magnitudes larger than $M_c$. As a complementary technique we analyzed the histograms of the hourly distribution. Histograms for all earthquakes at the North (Fig 13b) and the South (Fig. 13d) show a trend to accumulate events during night hours coinciding when the anthropogenic noise is lower. However, if the histograms are computed only for earthquakes with magnitude larger or equal to $M_c$, this trend almost disappears.

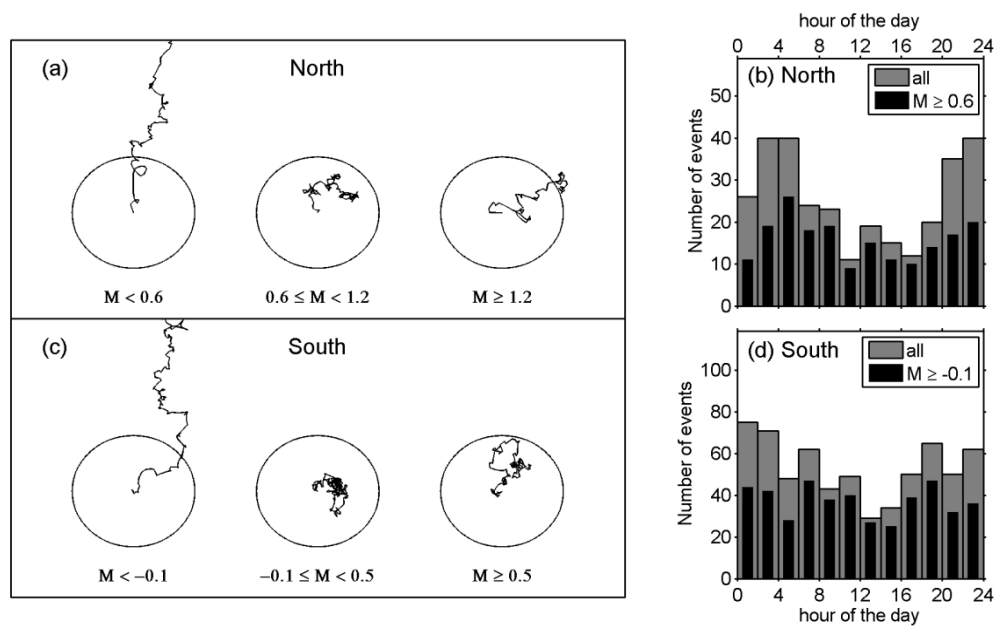

**Figure 13.** Rydelek and Sacks's test for the data of the improved catalogue split in North events (a) and South events (c). Each test shows the magnitude range. The circles represent the 95% confidence level for an equivalent random walk. Histograms for hourly distribution of the earthquakes belonging to North families (b) and South families (d). Gray histograms correspond to the total data set and black histograms to those events with magnitude larger than $M_c$.

The b-values obtained from the Gutenberg-Richter curves were 0.95 for North events (Fig. 12a) and 1.3 for South events (Fig. 12b). Since the number of events is high enough we have also studied the temporal evolution of this parameter. For that purpose, we have considered the maximum-likelihood method (Aki, 1965; Utsu, 1965; Woessner and Wiemer, 2005)

$$b = \frac{\log_{10}(e)}{\langle M \rangle - \left(M_c - \Delta M_{bin}/2\right)} \quad (6)$$

In this equation $\langle M \rangle$ is the averaged magnitude of the sample, $M_c$ the completeness magnitude computed before and $\Delta M_{bin}$ is the binning size of the magnitude catalogue. The uncertainty of the b-value computed by Eq. 6 was found by Shi and Bolt (1982):

$$\delta b = 2.30 \, b^2 \sqrt{\frac{\sum_{i=1}^{n}(M_i - \langle M \rangle)^2}{n(n-1)}} \quad (7)$$

Here $n$ is the size of the sample. To obtain a good resolution of the temporal evolution, b-values were computed with Equation 6 applied to samples of 40 events and with a step of 1 event. Figure 14 shows the temporal evolution of the b-value for North and South events. The North b-value is always lower than 1.0, starting with 0.6 on May and growing intermittently until September. The South b-values remain at 0.8 until September when they start to rise up to 1.6 and decrease again to 0.8 from October to December of 2004, with almost the same value continuing during 2005. These South b-values also experiment a slightly increase in April 2004. The uncertainty found for b-values is in all cases between 10 and 15%.

## 7. Discussion

### 7.1 Results interpretation

We analyzed the volcano-tectonic seismicity during the presumable volcanic reactivation of Tenerife in 2004. We applied different techniques in order to improve the seismic catalogue for better knowledge on the seismicity produced during the crisis.

Earthquakes located inland have been classified by means of cross-correlation of the seismic signals resulting in 60% of the events being grouped into only 6 families. Since the classification was done taking into account readings from different stations and thus, comparing the seismograms of each earthquake from different raypath orientation, it is reasonable that earthquakes of each family share a focal mechanism and a close hypocentral location.

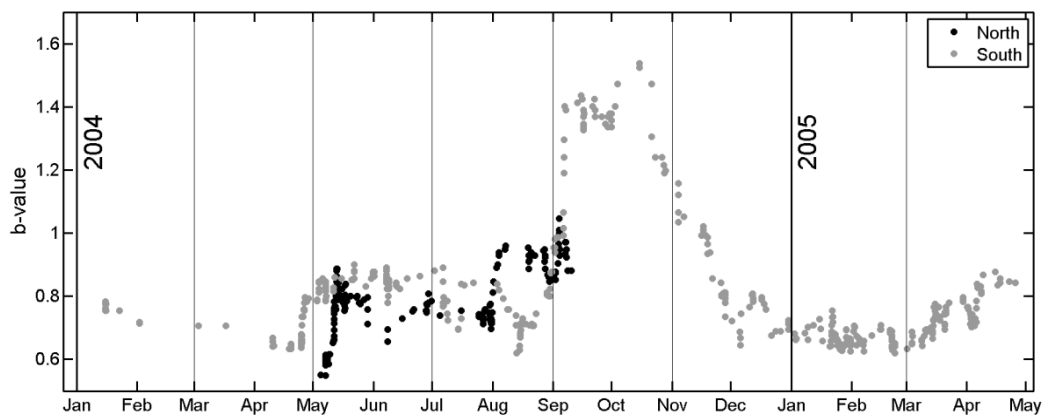

**Figure 14.** Temporal evolution of b-value computed with the maximum likelihood method with a window of 40 events and a step of one event. Curves have been computed separately for North events (black dots) and for South events (gray dots).

Earthquake families have been observed in many volcanoes like Decepcion (Ibáñez et al., 2003), Soufrière Hills (Rowe et al., 2004) or Nevado del Ruiz (Londomo and Sudo, 2001). In most cases these families are swarms lasting a few days, however, there are also examples of longer lasting swarms, with some even lasting several months as we observed in Tenerife, and in Merapi (Poupinet et al., 1996). Okada et al. (1981) already studied families in Usu volcano, and explained them in terms of barriers of different sizes and strengths. In that case the seismicity was produced by the deformation of a growing dome.

Hypocentral locations have been improved with a relative location algorithm (hypoDD). As a result of this methodology, hypocenters are better clustered and a depth has been assigned to each, which was almost impossible to do with an absolute location algorithm due to the low number of seismic stations. The relocation shows a tighter clustering of the families giving some self consistency to the results of both techniques (classification and relocation).

The 200 earthquakes reported by the IGN during 2004-2005 were just the tip of the iceberg. After reviewing the seismic signals from the station best located to study these anomalies (CCAN), we found about 3600 events in the same time period. We correlated the waveforms of the extra events with the average family waveform to define linked events and complete the catalogue. With this technique we could assign 800 events; in other words, we increased the number of earthquakes in the catalogue 5 times. However, it is noteworthy that only 33% of the extra events were assigned to families. We tried to classify the remaining events but did not obtain families large enough to be considered.

We computed the magnitude of the events by comparing the seismic signal levels. This method made it possible to obtain magnitudes for events detected only at the CCAN station and helped to homogenize the magnitudes of events in the IGN catalogue. Even if magnitudes are not exact, as this is not the best technique to obtain them, the analogy with the RMS should homogenize the catalogue, therefore; results obtained in the evolution of energy, and b-value should be trustworthy.

We have discovered the existence of two well defined seismogenic zones, one located NW of the Teide-PicoViejo complex and the other located on the SW border of Las Cañadas caldera. This discovery is one of the most important results of this work and gives us new keys for the interpretation of the seismicity. The following characteristics have been found:

(1) The centroids of the two zones are separated by more than 10 km, and are located at opposite sides of the central edifice of the volcano. Both zones are located at different depths and have incoherent temporal distribution, and thus, they can be considered to be produced by isolated faulty structures, tectonically unconnected.

(2) The mean depth of the North zone is around 7 km below sea level which corresponds approximately to the bottom of the basaltic shield of the island (Dañobeitia and Canales, 2000), while the South zone is located at a depth of about 4 km.

(3) The earthquake distribution of the North zone shows a structure dipping to NE. However, there can be doubts on the real existence of such geological structure due to the poor seismic network, mostly because families of earthquakes should be originated in small areas not larger than one km.

(4) Although previous studies had set the start of the seismic crisis in April of 2004, about 40 shocks in the South cluster were detected by CCAN during January 2004. Another interesting observation is the alternation of activity between the South and North zones which clearly points to an internal link between them. This fact refutes the theory of a gradual migration from North to South (Almendros et al., 2007; Martí et al., 2009).

(5) The magnitude-frequency distribution of the two seismogenic zones shows lower magnitudes for South events than for North events. More than 95% of the seismic energy was released by the North zone.

(6) The b-value remains almost stable for North events during the crisis, while for South events it suffered important variation reaching values of 1.6 in September 2004 when North activity disappeared. Finally, it returned to its normal value at the end of November. This evolution differs from that found by Martí et al. (2009) which showed a value of $b$ close to one until September 2004, and then it rose up to roughly two and remained there in 2005. These discrepancies may be due to the different data considered; they took into account earthquakes of the IGN catalogue in a radius of 25 km around Teide, while we have completed the database and took only those from local seismic zones.

The identification of these two seismic regions of alternating activity gives a new ingredient to the model proposed to explain observations related to the volcanic reactivation. There are many examples of volcanic regions with various differentiated seismogenic zones, like Pinatubo (Harlow et al., 1996), Popocatepetl (Lermo-Samaniego et al., 2006; De la Cruz-Reyna et al., 2008) or Merapi (Ratdomopurbo and Poupinet, 2000). However, it is unusual to find alternating activity between the different regions.

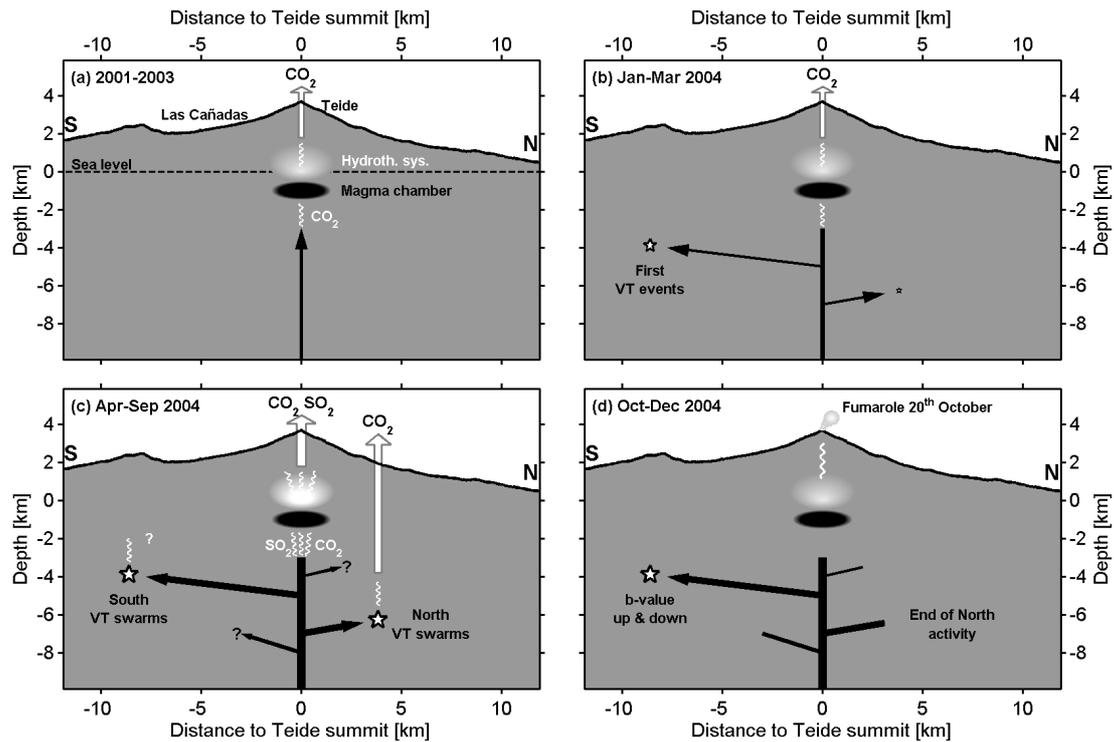

**Figure 15.** Scheme of the proposed model in four different stages. Black arrows display magma intrusions, white stars correspond to VT swarms and white arrows and lines show degassing. (a) Between 2001 and 2003 a deep magma intrusion or accumulation generates gas ascension affecting the hydrothermal system and increasing the diffuse $CO_2$ emission observed in Teide summit. (b) In January 2004 first VT swarms produced by the injection of magma. (c) From April to June 2004 the maximum magma intrusion produced degassing of $SO_2$ and VT swarms in North and South region. In the NW rift diffuse $CO_2$ increase was also observed. d) In October 2004 the North activity stopped and the South seismicity presents variation on the b-value coinciding with an increase of fumarolic activity on Teide summit.

The catalogue enhancement also helped us to discard some proposed explanations for the increased seismicity of the 2004 crisis in Tenerife. One of the main arguments of Carracedo and Troll (2006) refuting a volcanic origin of the anomalous seismicity was to associate it with intensive extraction of groundwater in these areas. However, we found arguments against such origin: First, seismicity related to changes in groundwater levels would reflect a seasonal component (e.g., Saar and Manga, 2003; Christiansen et al., 2006) which has not been observed. Ten years of data from CCAN failed to show rises in seismicity associated with refills of the aquifer due to rainfall, nor with water extraction. The only period of increased seismicity was seen during the crisis. Secondly, seismicity was located in two different zones which correspond to independent aquifer areas (Lillo and Márquez, 2007; Consejo Insular de Aguas, 2010). Lastly, the earthquakes were located excessively deep (most of them more than 2 km below sea level) to be directly related to aquifers located 1000 m above sea level.

Another model to explain the volcanic reactivation was proposed by Almendros et al. (2007). They suggested a deep magma injection at the beginning of April located under the northwest flank of Teide volcano and related to a basaltic magma chamber at the depth where they detected the first earthquakes (~15 km). This seismicity allowed the magma degassing which produced deep LP activity. The gas flow front reached the aquifer of Las Cañadas in the second half of May producing the tremor detected by their seismic arrays and at the time that they observed a dispersion of the volcano-tectonic earthquakes towards the southeast. This model was based in a migration of the seismicity that has not been observed in our new catalogue, and did not explain the activity of the southern seismogenic zone.

### 7.2 Tree-like intrusion model

We propose a new model to explain the results obtained by this work and which will also concur with different geophysical and geochemical observations reported in the literature. The model can be detailed in four different stages. Because the first evidence of a possible volcanic reactivation was reported to start in 2001 (Pérez et al., 2005), we start our model at that time.

(1) From 2001 to 2003 (Fig. 15a) some small magmatic intrusions affected the hydrothermal system increasing the $CO_2$ and $H_2S$ emission at the Teide summit (Pérez et al., 2005; Pérez and Hernández, 2007). These intrusions did not produce perceptible seismicity at any seismic station. Another possibility for the origin of these emissions could have been the degassing of the accumulating magma in a deep reservoir.

(2) During January 2004 (Fig. 15b) new magmatic injections arrived at the North and South seismogenic regions observed in this work. This is supported by the first seismicity observed at this time, mainly in the South (Fig. 10), however, the presence of small earthquakes of North families indicate that the magma could also have reached this region. The apparent simultaneity of activity on the two zones, together with the presence of magmatic gases at the summit of Teide, seems to indicate a common origin in a main intrusion located approximately below Teide instead of two independent intrusions. Fumaroles should be associated with the magma ascension below the hydrothermal system. The situation of the North zone suggests its relation with the NW rift and the position of the South zone can be correlated with the prolongation of the NE rift. According to this fact, magmatic injections could have been driven from the center in the direction of the two rifts, where fractures facilitated the magma ascension. The lack of activity from the end of January to the beginning of April could be related to a pause in the magma ascension.

(3) Between April and June 2004 (Fig. 15c) major magmatic intrusions occurred with a peak of the seismic activity. At that time the presence of magmatic gases like $SO_2$ in the diffuse emissions of the Teide summit reached the highest values (Pérez et al., 2005). During April and May the presence of diffuse $CO_2$ in the NW rift increased by a factor of 5 from previous values (Galindo 2005). No measurement of $CO_2$ was done in the south region at that time and thus there is no evidence on the increment of gases in this zone. It is possible that the magma injections were produced simultaneously, however, the seismicity of the two regions had a time lag of some days. To justify such behavior we must consider how the seismicity was originated. Rubin and Gillard (1998) explained the origin of low magnitude volcano-tectonic earthquakes with a dike-induced model. They suggested that the earthquakes could have originated in existing fractures, located adjacent to the tip of the dike or even at a certain distance. The propagation of the dike may produce the needed pressure to achieve the rupture of these fractures. In our model we believe that the dike intrusion has affected two clear zones of well defined fractures. Many seismic swarms started with a relatively high magnitude event, which suggests that the stress was accumulated over some time until the energy was released. The magnitude of the earthquakes in the South zone is lower than in those of the North zone, therefore, the North seismogenic structure is composed by faults with a larger capacity to store energy than that of the South, and thus the North zone requires higher pressures to reach rock failure. In consequence, if considering a constant increase of pressure in both regions, seismicity should appear first in the South region. Then, the difference in fault strength is a possible explanation for the apparent alternation of activity between the zones.

(4) At the end of September 2004 (Fig. 15d) the North seismic activity decreased drastically, which can be attributed to a cease of the magmatic ascension through the NW rift. Another possibility is that the magma intrusion at this rift was no longer affecting the seismogenic zone; however, the decrease in the diffuse $CO_2$ emissions in the region suggests at least a drop of the magmatic activity. At the same time, seismic activity in the South region showed some changes with a rise in the b-value (equivalent to a decay in the material stress) reaching its maximum value during the last weeks of October 2004. On the morning of 20 October 2004 an important increase in the fumarolic activity was observed at Teide's summit (Martí et al., 2009). The increasing b-value can be interpreted as the presence of magma or gases in the seismogenic zone, which decreases the stress of the materials. The release of gases could relax overpressure in the system, stopping the ascension of magma and resulting in the drop of the $b$-value which declined to its common level within two months.

There are two possible origins for the magmatic intrusion: 1) Local tectonic movements could generate the appearance of new fractures which facilitate the ascension of new magma as was proposed by Galindo (2005). 2) Overpressure in the underplating produced by ascending magma from the mantle could also produce ascension of magma. However, with the data available, we do not have enough information to choose between these possibilities.

It is noteworthy that most of the detected seismicity was located in two independent regions while there was no seismicity tracing the path of the magma intrusions. As we already pointed out, during the studied period (2004 and 2005) there were at least 1400 events without information about location or magnitude. Possibly these earthquakes were VT events produced by ascending magma, but too weak to be located and impossible to associate with any family. Other volcanic seismic signals like tremors or LP observed during this period (e.g., Tarraga et al. 2006; Almendros et al 2007; Carniel et al. 2008a) may be related to magma migration.

Despite the fact that the presence of earthquake swarms may not be an indicator of magma ascension there is another observation that supports this hypothesis: the gravimetric anomalies observed by Gottsmann et al. (2006). They interpreted the presence of positive anomalies from May 2004 to June 2005 as an increment on the density by a sub-surface mass addition. The position of the anomalies matches the location of the two seismogenic zones.

There are reasons to believe that the magmatic intrusion affected the central part of the island (e.g., the presence of $SO_2$) and could possibly contribute to the heating of the phonolitic magma chamber (Martí et al., 2009). The relation between basaltic intrusions and reactivation of phonolitic eruptions was already pointed out by Martí, et al. (2008) and should be taken into account in future seismo-volcanic reactivations, even if they do not end in basaltic eruption, like the case of 2004.

There is an historical example which could be produced by a model similar to the one proposed in this work. At the beginning of 1705 a series of eruptions occurred in the NE rift (Siete Fuentes, Fasnia and Arafo) and approximately a year later, in May 1706, a new volcano appeared in the NW rift (Garachico) separated by more than 25 km from the eruptive vents of the previous year. Despite the spatial separation of the eruptions, the short time lag suggests the occurrence of a single eruptive process; however, there is no petrological study to confirm this hypothesis. As in our model, the magma ascension could have been driven through the two rifts from a common source, but in that case the magma ascension ended in volcanic eruptions. This could be the usual behavior of the ascending magmas in Tenerife even in cases with eruptions in only one rift, but this will only be known in future reactivations of the island.

## 8. Conclusions

Different techniques were applied to successfully improve the seismic catalogue of volcano-tectonic events detected during the 2004 volcanic crisis of Tenerife. Seismicity can be classified into a small number of families by means of cross-correlation analysis. Using relative location algorithm we have been able to substantially improve the precision of hypocentral locations. This relocation has shown the presence of two seismogenic zones separated by more than 10 km and located at different depths. Thanks to the assignment performed using cross-correlation of the seismic signal compared with the obtained families, we increased the number of catalogued events from 200 to more than 900. The new catalogue shows a homogeneous detection level for the seismicity, which is a major improvement over the problems of completeness in the original seismic catalogue of the IGN. Finally, all magnitudes have been calculated for the new events and corrected for the catalogue events by a signal comparison method in order to homogenize the database of the study. Independent analysis of the two seismogenic regions shows different time evolution of the b parameter of the Gutenberg-Richter law.

We have shown how these techniques provide potentially powerful tools to substantially increase the precision of earthquake location and to add new events to the catalogue. In future works we propose to create a semiautomatic system able to improve the catalogue in real time which may evolve into an important procedure in the volcano monitoring system of Tenerife or in any other active volcano.

The model proposed herein to explain the volcanic reactivation of Tenerife is possibly the simplest one which will find accord between all results of the present study and those observations reported in previous works; although some details will possibly never be understood, like the origin of the magma ascension. We suggest a main intrusion from which magma rises in the direction of the two rifts and reaches the seismogenic zones where the magma pressure triggers the observed seismicity. The presence of magmatic gases in the summit of Teide is further evidence of this main intrusion affecting the central part of the island. In this case, such volcanic process may contribute to the heating of a shallow magma chamber, as already proposed by the study of previous phonolitic eruptions, which is something that should be taken into account in the study of future alterations in the volcanic activity of the island.

## Acknowledgements

We wish to thank Dr. Joan Martí for his suggestions and helpful discussions on the ascension model and to Kathy Martin for revising the English language which substantially improved the readability of the manuscript. We are also grateful to our colleagues in our institutions for their comments and kind support, especially to María José Blanco, Carmen López, Stavros Meletlidis, Pedro A. Torres and Almudena Gomis. R. Arámbula-Mendoza is acknowledged for his suggestions in the initial stage of this work. We also want to thank the constructive comments made by the two anonymous reviewers which have increased the quality of this paper. This work was supported by the Instituto Geográfico Nacional under the Spanish Ministerio de Fomento.